\documentclass[reprint,longbibliography,prb,amsmath,amssymb,amsfont]{revtex4-1} 
\usepackage[colorlinks, allcolors=blue]{hyperref}
\usepackage{graphicx} 
\usepackage{enumitem}
\usepackage{eurosym}
\usepackage[cal=dutchcal]{mathalfa}
\usepackage{csquotes}
\usepackage{mathtools}

\newcommand*\diff{\mathop{}\!\mathrm{d}}

\begin{document}
	\author{D. Lairez}
	\email{didier.lairez@polytechnique.edu}
	\affiliation{Laboratoire des solides irradi\'es, \'Ecole polytechnique,  CEA, CNRS, IPP,
		91128 Palaiseau, France}
	\title{A short derivation of Boltzmann distribution and Gibbs entropy formula\\ from the fundamental postulate}
	\date{\today}

\begin{abstract}
Introducing the Boltzmann distribution very early in a statistical thermodynamics course (in the spirit of Feynmann) has many didactic advantages, in particular that of easily deriving the Gibbs entropy formula. In this note, a short derivation is proposed from the fundamental postulate of statistical mechanics and basics calculations accessible to undergraduate students.
\end{abstract}
	
\maketitle

Clausius entropy, i.e. the state variable used in thermodynamics to account for heat exchanged, and Gibbs entropy, i.e. the one computed from the probability distribution $p_i$ of microstates, are the same physical quantity.
However, this link and the famous formula\footnote{In this note $S$ has no dimension and the temperature $T$ is in Joule, this amounts to incorporate the Boltzmann constant $k$ into $T$ rather than in $S$ for a gain of concision} 
\begin{equation}\label{GS}
	S=-\sum_i p_i\ln(p_i),
\end{equation} 
seem to be pulled out of a hat in most textbooks and undergraduate university courses.
These can be classified into categories lying between two main extreme approaches\,:
\begin{enumerate}
\item Those who start by defining statistical entropy with Eq.\,\ref{GS} (either without further justification, e.g. in Huang\cite{Huang_1987} or Reichl\cite{Reichl_2016}, or by introducing the formula \textit{via} information theory, e.g. in Balian\cite{Balian_1991} or Ben-Naim and Casadei\cite{Ben-Naim_2016})
and then show that from a \textquote{maximum entropy principle} this makes it possible to recover all of classical thermodynamics.
\item Those who first derive the Boltzmann probability distribution (or canonical distribution) of a thermalized system, then compute some statistical quantities, which by identification with classical equations of thermodynamics, lead to the statistical entropy formula (e.g. in Tien and Lienhard\cite{Tien_1979} or Sekerka\cite{Sekerka_2015}).
\end{enumerate}
Intermediate approaches can be found, for instance in Reif\cite{Reif_1965}, Chandler\, \cite{Chandler_1987} or Pathria\cite{Pathria_1996}, which basically start by defining (arbitrarily) the entropy of the microcanonical ensemble (Boltzmann entropy), then deriving the canonical distribution from it and finally join the above second category.

In my opinion, the magical side and the arbitrariness of the first and intermediate approaches can be disturbing for students, as is for many the connection between energy and  information (i.e. a number of possibilities offered to the system).
For this reason, the last category could be my favorite, but the main didactic difficulty lies in the derivation of the Boltzmann distribution from fundamental assumptions that would be easily stated and acceptable for all.

The classic way to derive the Boltzmann distribution\cite{Tien_1979,Sekerka_2015} is by seeking which distribution is the most probable, that is to say which distribution allows the multiplicity of a macrostate to be maximum. This classic method presents some didactic difficulties. Firstly, it involves Lagrange multipliers that is usually not introduced in undergraduate courses. Secondly, it uses also the Stirling approximation, which derivation is far from being straightforward. Lastly, it is not the multiplicity but its logarithm which is maximized, and even if it is the same thing it has a slight taste of arbitrariness and gives the feeling that we know the solution in advance.

In this note we propose a short derivation of the Boltzmann distribution which avoid these issues and starts from the fundamental postulate of statistical mechanics. So very simply, the Gibbs formula can be obtained without suffering \textit{via} the free energy and the partition function.

\section{Free energy}

Let us consider a system at temperature $T$, denote $U$ the internal energy, $S$ the entropy and $W$ the work exchanged with the surroundings. For any process that the system can undergo, the Clausius inequality writes\,:
\begin{equation}
	\diff U -T\diff S\le W
\end{equation}
For processes occurring at constant temperature, this can be rewritten as
\begin{equation}
	\diff (U -TS)\le W
\end{equation}
The differential $	\diff (U -TS)$ introduces a thermodynamical potential\cite{Callen_1985} named free energy (or Helmholtz free energy):
\begin{equation}\label{freeEnergy1}
	F=U-TS
\end{equation}
As $S=-\partial F/\partial T$, the later equation can be rewritten as
\begin{equation}\label{Gibbs-Helmholtz}
	\begin{array}{rl}
		U&=F+TS \displaystyle =F-T\frac{\partial F}{\partial T}  =F+\frac{1}{T}\frac{\partial F}{\partial (1/T)}\\ \\
		&\displaystyle =\frac{\partial(F/T),}{\partial (1/T)}
	\end{array}
\end{equation}
This equation is one of the Gibbs-Helmholtz relations.

\section{Fundamental postulate}\label{fp}

Imagine a large isolated system, with constant total energy $E$, divided into a large number $N$ of subparts with number $n=1\cdots N$.
\begin{equation}
	E=\sum_{n=1}^N E_n=\text{cst},
\end{equation}
Let us define a microstate by the multiplet  obtained by the energies of subparts\,:$$\textrm{microstate} = (E_1, E_2,\cdots,E_N),$$
and denote $W_N$ the number of possible microstates the system can adopt.

A finite resolution for the energy and the existence of a upper boundary, allow us to discretize the possible energy levels of subparts into a set of $w$ values\,: 
\begin{equation}
E_n \in \{\epsilon_0, \epsilon_1, \cdots \epsilon_{w-1}\} \quad \textrm{with} \quad \epsilon_i=i\epsilon
\end{equation}
where $\epsilon$ is the smallest observable energy exchange.

The system is dynamical and at every time an elementary transition can occur from a given microstate to another\,:
$$\begin{array}{c}
(E_1,\cdots, E_k,\cdots,E_{l},\cdots,E_N) \\ \\
\text{elementary transition} \displaystyle\left\updownarrow\vphantom{\int}\right. \phantom{\text{elementary transition}}\\ \\
 (E_1,\cdots, E_k-\epsilon,\cdots,E_{l}+\epsilon,\cdots,E_N)
\end{array}
$$
Although elementary transitions are fundamentally deterministic, the finite resolution at which the initial microstate is known prevent us to predict the final one.
The microstate in which the system can be found is thus a random variable.

Based on the information we have about microstates, there is no reason to believe that one is more likely than another. The fundamental postulate of statistical mechanics, which is a variation of the Laplace's \textquote{principle of insufficient reason}, states that\,:
at the equilibrium all the $W_N$ microstates of an isolated system have equal probability $1/W_N$.

\section{Boltzmann distribution}

The energy level of a given subpart (a closed system) fluctuates due to elementary  transitions. 
To simplify the notation, let us note the probability $\mathbb{P} (E_n=\epsilon_i)$ that a given subpart $n$ has the energy level $\epsilon_i$ as\,:
$$
\mathbb{P} (E_n=\epsilon_i)=p_n(\epsilon_i),
$$

Consider two given subparts $(n, m)$ sufficiently far the one from the other so that they are uncoupled\,: $E_n$ and $E_m$ are independent random variables. This is possible since $N$ is very large. Denote
\begin{equation}
\mathcal E=E_n+E_m	
\end{equation}
The probability for $\mathcal E$ to have a given value $x$ is\,:
\begin{equation}\label{sume}
	\mathbb{P} (\mathcal E = x) = \displaystyle\sum_{i=0}^{w-1} p_n(\epsilon_i)p_m(x -\epsilon_i)
\end{equation}
Once given $(n,m)$ and $\mathcal E=x$, the rest of the $N-2$ subparts have a fixed energy $E-x$ so that it can be viewed as isolated with $W_{N-2}$ equiprobable microstates (in virtue of the fundamental postulate) that only depend on $E-x$ but not on
the peculiar state the subparts $n$ and $m$ adopt. In Eq.\ref{sume}, the three events\,: 1)~subpart $n$ has energy $\epsilon_i$; 2)~subpart $m$ has energy $x-\epsilon_i$; and 3)~the rest adopts any particular microstate (with energy $E-x$); are complementary events for any microstate of the whole (with probability $1/W_N$). So that one can always write\,:
\begin{equation}
	\frac{1}{W_N} = p_n(\epsilon_i)p_m(x -\epsilon_i)\frac{1}{W_{N-2}}
\end{equation}
It follows that in Eq.\ref{sume}, the $w$ terms $p_n(\epsilon_i)p_m(x -\epsilon_i)$
of the sum are all the same and equal to $W_{N-2}/W_N$\,:
\begin{equation}\label{sume1}
\mathbb{P} (\mathcal E=x) = w p_n(\epsilon_i)p_m(x -\epsilon_i),\quad \forall i\in\{0,\cdots w-1\}
\end{equation}
For $\epsilon_i=0$ or for $\epsilon_i=x$, one gets respectively\,:
\begin{align}
	\mathbb{P} (\mathcal E=x) = w p_n(0)p_m(x)\\
		\mathbb{P} (\mathcal E=x)	= w p_n(x)p_m(0)
\end{align}
From the ratio of these two equations, it follows that whatever $x$, $p_n(x)$ and $p_m(x)$ are proportional.
If these two probability distributions are normalized, then they are identical\,:
\begin{equation}
	p_n = p_m= p
\end{equation}
Therefore whatever $a$ and $b$ in $\{0,\cdots w-1\}$, one can write
\begin{equation}\label{sume3}
	\mathbb{P} (\mathcal E=\epsilon_a+\epsilon_b) = w p(\epsilon_a)p(\epsilon_b) =w p(0)p(\epsilon_a+\epsilon_b)
\end{equation}
Thus
$
p(\epsilon_a) p(\epsilon_b)= p(0) p(\epsilon_a+\epsilon_b)	
$ whatever $\epsilon_a$ and $\epsilon_b$,
which is a definition for the exponential function, that can be written as\,:
\begin{equation}
p(\epsilon_i)\propto e^{-\beta \epsilon_i}	
\end{equation}
where $\beta$ is a constant that is currently undetermined.
As the energy levels are discrete, this exponential distribution can be rewritten as\,:
\begin{equation}\label{pi}
	p_i=\frac{1}{Z}e^{-\beta \epsilon_i} 
	\quad\textrm{with}\quad
	Z=\sum_{i=0}^{w-1} e^{-\beta \epsilon_i}
\end{equation}
This is the Boltzmann or canonical distribution (provided $\beta$ is elucidated) with $Z$ the partition function. 
The physical meaning of the constant $\beta$ is given in the next section directly by making the link with thermodynamics and entropy.

\section{Gibbs entropy}\label{GibbsEntropy}

From the probability distribution of energy levels, some additive state-quantities of a closed system can be derived. The internal energy is the mathematical expectation of its energy level\,:
\begin{equation}\label{intU}
	U = \sum_i p_i \epsilon_i
\end{equation}
Let us compute the derivative of $\ln(Z)$ versus $1/T$\,:
\begin{equation}
\begin{array}{rl}
	\displaystyle\frac{\diff(\ln(Z))}{\diff (1/T)} 
	& =\displaystyle \frac{1}{Z} \frac{\diff Z}{\diff(1/T)} =\displaystyle  \frac{1}{Z}  \sum_{i=0}^{w-1}-\epsilon_i e^{-\beta\epsilon_i}=\displaystyle - \sum_{i=0}^{w-1}p_i\epsilon_i \\ \\
	&= -U
\end{array}	
\end{equation}
Identification with Eq.\ref{Gibbs-Helmholtz} gives the free energy:
\begin{equation}\label{Zfree}
	F = - T \ln(Z)
\end{equation}
This allows us to rewrite Eq.\ref{pi} as
$
	p_i=e^{F/T-\beta\epsilon_i}
$
giving\,:
\begin{equation}\label{pi2}
	\epsilon_i= \frac{1}{\beta T}  \left[{F -T\ln(p_i)}\right]
\end{equation}
By using Eq.\ref{intU}, one gets\,:
\begin{equation}\label{Umechstat}
	\begin{array}{rl}
		U  &\displaystyle = \frac{1}{\beta T} \sum_{i=0}^{w-1} p_i \left[{F -T\ln(p_i)}\right]\\ \\
		&\displaystyle =\frac{1}{\beta T} \left[{F -T\sum_{i=0}^{w-1} p_i\ln(p_i)} \right]
	\end{array}	
\end{equation}
From Eq.\ref{freeEnergy1} one has
\begin{equation}\label{UfromF}
	U = F + TS,
\end{equation}
so that identifying Eq.\ref{Umechstat} with Eq.\ref{UfromF} gives\,:
\begin{flalign}
	\beta &= \frac{1}{T}\\
	S &= - \sum_{i=0}^{w-1} p_i\ln(p_i) 
\end{flalign}
That is what we were looking for.

\section{Conclusion}

The Boltzmann distribution and the Gibbs entropy formula were derived from the sole assumption of the equiprobability of microstates in which an isolated system can be found, i.e. from the fundamental postulate of statistical mechanics. This postulate is quite acceptable for students. It suffices to convince them to think about the probability of the outcome of any game of chance. The mathematics of the derivation is without any difficulty for undergraduate student. This therefore makes it possible to introduce Boltzmann distribution and statistical entropy very early in a course in statistical thermodynamics without arbitrariness. This presents at least two advantages. Firstly,
it is in this way that things were done by the pioneers. Planck, Boltzmann, Gibbs were all thermodynamicists before founding statistical mechanics. They clearly had in mind the whole phenomenology that their statistics had to account for.
A different approach would pull the course towards a more abstract side. Secondly, the Boltzmann distribution is a fundamental tool that opens the door to many interesting examples that can illustrate the course and make it less abstract. In fact, the Boltzmann distribution is one of the most fundamental results of statistical mechanics. Feynmann
in his famous lectures, starts the statistical mechanics book\cite{Feynmann_1972} in page 1, 3rd line, by giving the Boltzmann distribution. I think that a short derivation would not harm the pedagogy of the whole.

\begin{acknowledgments}
	I thank Guy Amit (BIU) for his interesting remarks and suggestions.
\end{acknowledgments}

\bibliography{\string~/Documents/weri_biblio.bib}

\begin{thebibliography}{12}%
\makeatletter
\providecommand \@ifxundefined [1]{%
 \@ifx{#1\undefined}
}%
\providecommand \@ifnum [1]{%
 \ifnum #1\expandafter \@firstoftwo
 \else \expandafter \@secondoftwo
 \fi
}%
\providecommand \@ifx [1]{%
 \ifx #1\expandafter \@firstoftwo
 \else \expandafter \@secondoftwo
 \fi
}%
\providecommand \natexlab [1]{#1}%
\providecommand \enquote  [1]{``#1''}%
\providecommand \bibnamefont  [1]{#1}%
\providecommand \bibfnamefont [1]{#1}%
\providecommand \citenamefont [1]{#1}%
\providecommand \href@noop [0]{\@secondoftwo}%
\providecommand \href [0]{\begingroup \@sanitize@url \@href}%
\providecommand \@href[1]{\@@startlink{#1}\@@href}%
\providecommand \@@href[1]{\endgroup#1\@@endlink}%
\providecommand \@sanitize@url [0]{\catcode `\\12\catcode `\$12\catcode
  `\&12\catcode `\#12\catcode `\^12\catcode `\_12\catcode `\%12\relax}%
\providecommand \@@startlink[1]{}%
\providecommand \@@endlink[0]{}%
\providecommand \url  [0]{\begingroup\@sanitize@url \@url }%
\providecommand \@url [1]{\endgroup\@href {#1}{\urlprefix }}%
\providecommand \urlprefix  [0]{URL }%
\providecommand \Eprint [0]{\href }%
\providecommand \doibase [0]{http://dx.doi.org/}%
\providecommand \selectlanguage [0]{\@gobble}%
\providecommand \bibinfo  [0]{\@secondoftwo}%
\providecommand \bibfield  [0]{\@secondoftwo}%
\providecommand \translation [1]{[#1]}%
\providecommand \BibitemOpen [0]{}%
\providecommand \bibitemStop [0]{}%
\providecommand \bibitemNoStop [0]{.\EOS\space}%
\providecommand \EOS [0]{\spacefactor3000\relax}%
\providecommand \BibitemShut  [1]{\csname bibitem#1\endcsname}%
\let\auto@bib@innerbib\@empty
\bibitem [{Note1()}]{Note1}%
  \BibitemOpen
  \bibinfo {note} {In this paper $S$ has no dimension and the temperature $T$
  is in Joule, this amounts to incorporate the Boltzmann constant $k$ into $T$
  rather than in $S$ for a gain of concision}\BibitemShut {NoStop}%
\bibitem [{\citenamefont {Huang}(1991)}]{Huang_1987}%
  \BibitemOpen
  \bibfield  {author} {\bibinfo {author} {\bibfnamefont {K.}~\bibnamefont
  {Huang}},\ }\href
  {https://www.wiley.com/en-us/Statistical+Mechanics%2C+2nd+Edition-p-9780471815181}
  {\emph {\bibinfo {title} {{S}tatistical {M}echanics}}},\ \bibinfo {edition}
  {2nd}\ ed.\ (\bibinfo  {publisher} {J. Wiley \& sons},\ \bibinfo {year}
  {1991})\BibitemShut {NoStop}%
\bibitem [{\citenamefont {Reichl}(2016)}]{Reichl_2016}%
  \BibitemOpen
  \bibfield  {author} {\bibinfo {author} {\bibfnamefont {L.E.}\ \bibnamefont
  {Reichl}},\ }\href
  {https://www.wiley.com/en-dk/A+Modern+Course+in+Statistical+Physics,+4th+Edition-p-9783527413492}
  {\emph {\bibinfo {title} {A modern course in statistical physics}}}\
  (\bibinfo  {publisher} {Wiley},\ \bibinfo {year} {2016})\BibitemShut
  {NoStop}%
\bibitem [{\citenamefont {Balian}(1991)}]{Balian_1991}%
  \BibitemOpen
  \bibfield  {author} {\bibinfo {author} {\bibfnamefont {R.}~\bibnamefont
  {Balian}},\ }\href {\doibase 10.1007/978-3-540-45475-5} {\emph {\bibinfo
  {title} {From microphysics to macrophysics}}}\ (\bibinfo  {publisher}
  {Springer Berlin Heidelberg},\ \bibinfo {year} {1991})\BibitemShut {NoStop}%
\bibitem [{\citenamefont {Ben-Naim}\ and\ \citenamefont
  {Casadei}(2016)}]{Ben-Naim_2016}%
  \BibitemOpen
  \bibfield  {author} {\bibinfo {author} {\bibfnamefont {A.}~\bibnamefont
  {Ben-Naim}}\ and\ \bibinfo {author} {\bibfnamefont {D.}~\bibnamefont
  {Casadei}},\ }\href {\doibase 10.1142/10300} {\emph {\bibinfo {title} {Modern
  thermodynamics}}}\ (\bibinfo  {publisher} {World Scientific},\ \bibinfo
  {year} {2016})\BibitemShut {NoStop}%
\bibitem [{\citenamefont {Tien}\ and\ \citenamefont
  {Lienhard}(1979)}]{Tien_1979}%
  \BibitemOpen
  \bibfield  {author} {\bibinfo {author} {\bibfnamefont {C.~L.}\ \bibnamefont
  {Tien}}\ and\ \bibinfo {author} {\bibfnamefont {J.H.}\ \bibnamefont
  {Lienhard}},\ }\href@noop {} {\emph {\bibinfo {title} {Statistical
  thermodynamics}}}\ (\bibinfo  {publisher} {Hemisphere Pub. Corp.},\ \bibinfo
  {year} {1979})\BibitemShut {NoStop}%
\bibitem [{\citenamefont {Sekerka}(2015)}]{Sekerka_2015}%
  \BibitemOpen
  \bibfield  {author} {\bibinfo {author} {\bibfnamefont {R.~F.}\ \bibnamefont
  {Sekerka}},\ }\href@noop {} {\emph {\bibinfo {title} {Thermal physics}}}\
  (\bibinfo  {publisher} {Carnegie Mellon University},\ \bibinfo {year}
  {2015})\BibitemShut {NoStop}%
\bibitem [{\citenamefont {Reif}(1965)}]{Reif_1965}%
  \BibitemOpen
  \bibfield  {author} {\bibinfo {author} {\bibfnamefont {F.}~\bibnamefont
  {Reif}},\ }\href@noop {} {\emph {\bibinfo {title} {Fundamentals of
  Statistical and Thermal Physics}}}\ (\bibinfo  {publisher} {McGraw-Hill},\
  \bibinfo {year} {1965})\BibitemShut {NoStop}%
\bibitem [{\citenamefont {Chandler}(1987)}]{Chandler_1987}%
  \BibitemOpen
  \bibfield  {author} {\bibinfo {author} {\bibfnamefont {D.}~\bibnamefont
  {Chandler}},\ }\href@noop {} {\emph {\bibinfo {title} {Introduction to modern
  statistical mechanics}}}\ (\bibinfo  {publisher} {Oxford university press},\
  \bibinfo {year} {1987})\BibitemShut {NoStop}%
\bibitem [{\citenamefont {Pathria}(1996)}]{Pathria_1996}%
  \BibitemOpen
  \bibfield  {author} {\bibinfo {author} {\bibfnamefont {R.K.}\ \bibnamefont
  {Pathria}},\ }\href {\doibase 10.1016/B978-075062469-5/50000-1} {\emph
  {\bibinfo {title} {Statistical mechanics}}},\ \bibinfo {edition} {2nd}\ ed.\
  (\bibinfo  {publisher} {Butterworth Heinemann},\ \bibinfo {year}
  {1996})\BibitemShut {NoStop}%
\bibitem [{\citenamefont {Callen}(1985)}]{Callen_1985}%
  \BibitemOpen
  \bibfield  {author} {\bibinfo {author} {\bibfnamefont {H.~B.}\ \bibnamefont
  {Callen}},\ }\href@noop {} {\emph {\bibinfo {title} {Thermodynamics and an
  introduction to thermostatistics}}},\ \bibinfo {edition} {2nd}\ ed.\
  (\bibinfo  {publisher} {J. Wiley \& sons},\ \bibinfo {year}
  {1985})\BibitemShut {NoStop}%
\bibitem [{\citenamefont {Feynmann}(1972)}]{Feynmann_1972}%
  \BibitemOpen
  \bibfield  {author} {\bibinfo {author} {\bibfnamefont {R.P.}\ \bibnamefont
  {Feynmann}},\ }\href@noop {} {\emph {\bibinfo {title} {Statistical mechanics:
  a set of lectures}}}\ (\bibinfo  {publisher} {Westview Press},\ \bibinfo
  {year} {1972})\BibitemShut {NoStop}%
\end{thebibliography}%

\end{document}